# ABOUT ONE APPLICATION OF A COMPLEX VARIABLE FUNCTION TO INVESTIGATION OF INTERACTION OF A DISLOCATION WITH A PORE IN POWDER MATERIALS

A. Kniazeva (Yankevich)

*The Institute of Mechanics and Reliability of Machines
of the National Academy of Sciences of Belarus, Minsk, Republic of Belarus
e-mail: alena.kniazeva@tut.by*

*Interaction of a rectilinear screw dislocation with a parabolic pore is prospected.*

Improvement of reliability and efficiency of products are some of the most important goals of the present-day mechanical engineering. They are significant characteristics of competitiveness. Mass reduction due to application of the newest technologies is not less important task. Powder metallurgy plays an acknowledged role in an achievement of all these goals. Fabrication of details from powder materials creates good conditions for complete mechanization and automation of production, economy of resources and abatement of costs of mechanical treatment. It is very important to understand processes in powder materials to provide quality of final products.

Mechanical behavior of metal materials is under control of mobility of dislocations in fields of forces with a different nature and intensity. An additional kind of force acts on a dislocation in powder materials. It is called the force of dislocation "reflection". This force tends to push the dislocation on the pore surface. Determination of the stress distribution near the pore surface (under acting forces) is one of the most important aspects in describing of mechanical behavior of the powder materials. This problem is of extremely significance for understanding of mechanics of plastic deformation and failure of these materials. In this article it is solved using methods of the complex variable functions theory.

Let's consider a simplest configuration which consists of one pore (surface or matrix) and one dislocation. Pores are not atomic-sharp after compressing and mixing. That is why their surfaces will be approximated by parabolas. Let an origin of a coordinate system is situated in the parabola pole. Pick of the parabola is at point (P/2, 0) of the complex plane Z=X+iY. A screw dislocation is parallel to the third coordinate axis and goes through the point ($X_0$, 0), which belongs to the parabola symmetry line. Let's turn the problem of an interaction of the dislocation with the pore into the well-known problem of an interaction dislocation with a flat surface.

We can reflect conformally exterior of the pore with the screw dislocation into the half-plane Re W > 0 (W=U+iV) [1] (Fig. 1) by means of analytical complex variable function

$$w = \sqrt{z} - \sqrt{\frac{P}{2}}. \tag{1}$$

According to [2], complex potential of a dislocation field is described by the next equation provided that there is a free surface

$$f(W) = \frac{b}{2\cdot\pi} \ln \frac{W - U_0}{W + U_0}, \tag{2}$$

$b$ is a dislocation Burgers vector.

We obtain the complex potential of the field of the screw dislocation in the plane Z after the substitution (1) into the expression (2)

$$f(z) = \frac{b}{2\cdot\pi} \ln \frac{\left(\sqrt{Z} - \sqrt{P/2}\right) - \left(\sqrt{X_0} - \sqrt{P/2}\right)}{\left(\sqrt{Z} - \sqrt{P/2}\right) + \left(\sqrt{X_0} - \sqrt{P/2}\right)}. \tag{3}$$

Stress components are connected with complex potential throught the equation of the linear fracture mechanics [3]

$$\tau_{yz} + i\tau_{xz} = \mu f'(z).$$

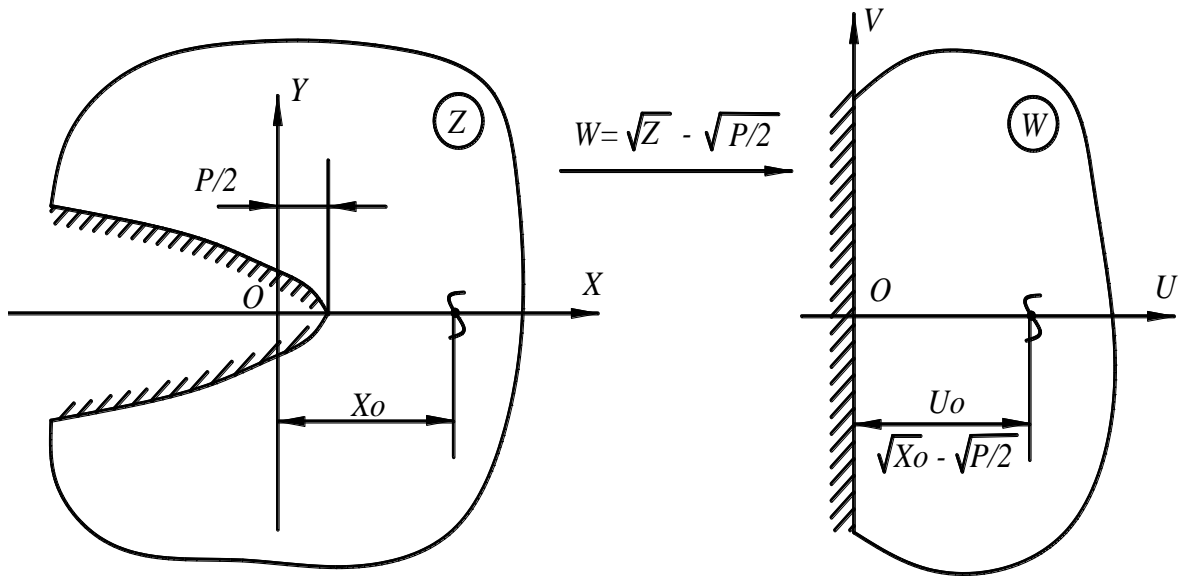

*Figure 1. Reflection of the exterior of the parabola $y^2 = 2p(x - p/2)$ into the half-plane $\operatorname{Re} w > 0$.*

The derivative of the complex potential (3) is calculated as

$$f'(z) = \frac{b}{2 \cdot \pi} \cdot \frac{\sqrt{X_0} - \sqrt{P/2}}{\left(\sqrt{Z} - \sqrt{X_0}\right)\left(\sqrt{Z} + \sqrt{X_0} - 2\sqrt{P/2}\right)\left(\sqrt{Z} - \sqrt{P/2}\right)} \quad (4)$$

Hence shear stress on the positive part of the real axis (into the material, as shown on the Figure 1) may be calculated from the next relationship

$$\tau_{yz} = \frac{\mu b}{2 \cdot \pi} \cdot \frac{\sqrt{X_0} - \sqrt{P/2}}{\left(\sqrt{X} - \sqrt{X_0}\right)\left(\sqrt{X} + \sqrt{X_0} - 2\sqrt{P/2}\right)\left(\sqrt{X} - \sqrt{P/2}\right)} \quad (5)$$

We can determine strain energy on the base of knowledge about the stress field near the dislocation. It is one of the most important characteristics of a system "dislocation – parabolic pore". This characteristic describes mobility of the dislocation and therefore controls properties of metal materials. Crystal lattice strain energy decreases when approaching to the pore surface. It happens because of the fact that the free surface doesn't produce stresses which may block dislocation displacement. We used a method of a calculation of a job, required to move the dislocation, to obtain strain energy of the screw dislocation in the parabolic pore presence [4]

$$E = 2\int_0^b db \int_{\sqrt{P/2}}^{(\sqrt{X_0}+\sqrt{P/2})+R_0} \tau_{yz} dx = \frac{\mu b^2}{4\pi}\left[\ln\frac{4}{R_0} + \ln\left(\sqrt{X_0}-\sqrt{P/2}\right)^2\right], \quad (6)$$

$R_0$ is a radius of a dislocation core.

In many problems formulation it is convenient to use concert of a force $F$, which acts on dislocation. This force is determined as energy gradient

$$F = -\frac{\partial E}{\partial X_0} = -\frac{\mu b^2}{4\pi}\frac{1}{\left(\sqrt{X_0}-\sqrt{P/2}\right)^2} \quad (7)$$

In the equation (7) $F$ is the force of force of dislocation "reflection", which attracts the dislocation to the free surface of the pore.

High local stresses near pores are main sources of decreasing of strength characteristics of construction steels in comparison with cast ones. Parabolic pores are characterized by small values of poles P/2 after the cold stamping. In such conditions we can use equations of linear fracture mechanics for the estimation of the stress concentration values near the pore peak. This approach is proved by experimental observations that peak of a sharp crack under external load has a parabolic shape [5]. Then the effective coefficient of stress concentration will be determined by the next equation at the pick of the parabolic pore

$$K_D = \lim_{x \to P/2}\left\{\sqrt{2\pi}\left(\sqrt{x}-\sqrt{P/2}\right)\tau_{yz}\right\} = -\frac{\mu b}{\sqrt{2\pi}}\frac{1}{\left(\sqrt{X_0}-\sqrt{P/2}\right)} \quad (8)$$

Equation (8) shows that in the absence of external loads the dislocation creates the stress concentration around the pick, which is typical for parabolic pore. The concentration value changes similarly to a function $(2\pi X_0)^{-1/2}$ of distance. Though the equation (8) is obtained for the configuration which is presented at the figure 1, it is also valid for pores (surface or matrix) with any shape. And coefficient $K_D$ is a modification of the coefficient $K_{1C}$ in the linear fracture mechanics. It includes parameters which take into account a deviation of the real pore from an ideal crack.

Let's then consider the case when screw dislocation is situated at an arbitrary point $Z_0$ of the complex plane Z. The complex potential of such configuration is described as follows

$$f(z) = \frac{b}{2\cdot\pi} \ln \frac{\left(\sqrt{Z}-\sqrt{P/2}\right)-\left(\sqrt{Z_0}-\sqrt{P/2}\right)e^{i\varphi/2}}{\left(\sqrt{Z}-\sqrt{P/2}\right)+\left(\sqrt{Z_0}-\sqrt{P/2}\right)e^{-i\varphi/2}} \quad (13)$$

The stress component $\tau_{yz}$ which acts along the real axis X will be the next

$$\tau_{yz} = \frac{\mu b}{2\cdot\pi\left(\sqrt{X}-\sqrt{\frac{P}{2}}\right)} \cdot \frac{\left[\left(\sqrt{X}-\sqrt{\frac{P}{2}}\right)^2 - \left(\sqrt{X_0}-\sqrt{\frac{P}{2}}\right)^2\right]\left(\sqrt{X_0}-\sqrt{\frac{P}{2}}\right)\cos(\varphi/2)}{\left[\left(\sqrt{X}-\sqrt{\frac{P}{2}}\right)^2 + \left(\sqrt{X_0}-\sqrt{\frac{P}{2}}\right)^2\right]^2 - 4\left(\sqrt{X}-\sqrt{\frac{P}{2}}\right)^2\left(\sqrt{X_0}-\sqrt{\frac{P}{2}}\right)^2 \cos^2(\varphi/2)}, \quad (14)$$

and a coefficient of stress intensity will be equal to

$$K_D = -\frac{\mu b}{\sqrt{2\pi}} \frac{1}{\left(\sqrt{X_0}-\sqrt{P/2}\right)} \cos\left(\frac{\varphi}{2}\right) \quad (15)$$

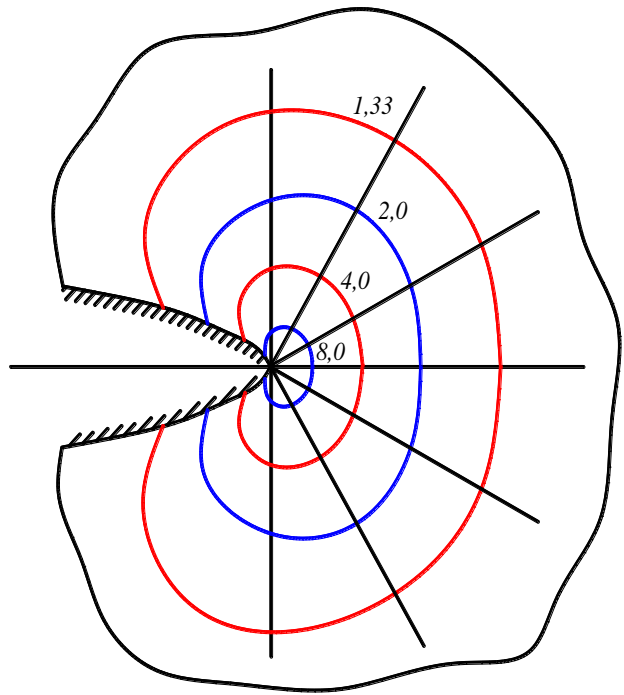

*Figure 2. Lines of the level of the effective concentration coefficient for the screw dislocation near the parabolic pore*

Lines of the level of the effective concentration coefficient $K_D$ are shown at the figure 2. They have shape of cardioid. A value of the dislocation radius vector decreases while increasing of the $K_D$ value, which is expressed in terms of dimensionless units.